\documentclass[utf8]{frontiersFPHY} % for Physics and Applied Mathematics and Statistics articles

\usepackage{url,hyperref,lineno,microtype,subcaption}
\usepackage[onehalfspacing]{setspace}
\usepackage{ulem}
\usepackage{dsfont}
\usepackage{bm}

\usepackage{cancel}
\def\keyFont{\fontsize{8}{11}\helveticabold }
\def\firstAuthorLast{Castro-Villarreal {et~al.}} %use et al only if is more than 1 author
\def\Authors{Pavel Castro-Villarreal\,$^{1,*}$, and J. E. Ram\'irez\,$^{2,*}$}
\def\Address{$^{1}$Facultad de Ciencias en F\'isica y Matem\'aticas, Universidad Aut\'onoma de Chiapas, Carretera Emiliano Zapata, Km. 8, Rancho San Francisco, 29050 Tuxtla Guti\'errez, Chiapas, M\'exico. \\
$^{2}$Centro de Agroecolog\'ia,
Instituto de Ciencias,
Benemérita Universidad Autónoma de Puebla, Apartado Postal 165, 72000 Puebla, Pue., M\'exico}
\def\corrAuthor{Correspondings Authors}
\def\corrEmail{$^{1}$
pcastrov@unach.mx,\\ 
$^{2}$jhony.ramirezcancino@viep.buap.mx}

\begin{document}
\onecolumn
\firstpage{1}

\title[Semiflexible polymer enclosed in a compact domain]{Semiflexible polymer enclosed in a 3D compact domain
} 

\author[\firstAuthorLast ]{\Authors} %This field will be automatically populated
\address{} %This field will be automatically populated
\correspondance{} %This field will be automatically populated

\extraAuth{}% If there are more than 1 corresponding author, comment this line and uncomment the next one.
%\extraAuth{corresponding Author2 \\ Laboratory X2, Institute X2, Department X2, Organization X2, Street X2, City X2 , State XX2 (only USA, Canada and Australia), Zip Code2, X2 Country X2, email2@uni2.edu}

\maketitle

\begin{abstract}

%%% Leave the Abstract empty if your article does not require one, please see the Summary Table for full details.
\section{}
The conformational states of a semiflexible polymer enclosed in a volume $V:=\ell^{3}$ are studied as stochastic realizations of paths using the stochastic curvature approach developed in [Rev. E 100, 012503 (2019)], in the regime whenever $3\ell/\ell_ {p}> 1$, where $\ell_{p}$ is the persistence length. The cases of a semiflexible polymer enclosed in a cube and sphere are considered. In these cases, we explore the Spakowitz-Wang type polymer shape transition, where the critical persistence length distinguishes between an oscillating and a monotonic phase  at the level of the mean-square end-to-end distance. This shape transition provides evidence of a universal signature of the behavior of a semiflexible polymer confined in a compact domain.

\tiny
 \keyFont{ \section{Keywords:} semiflexible polymer, stochastic curvature, shape transition, critical persistence length, mean square end-to-end distance.} %All article types: you may provide up to 8 keywords; at least 5 are mandatory.
\end{abstract}

\section{Introduction}

Semiflexible polymers is a term coined to understand a variety of physical systems that involve linear molecules. The most popular polymers are industrial plastics, like polyethylene or polystyrene, with various applications in daily life \cite{RONCA2017247, polystyrene}.
Another prominent example is the DNA compacted in the nucleus of cells or viral DNA/RNA packed in capsids \cite{Fal-Cifra2010, Fal-Locker2006}. These last examples are of particular interest since they are confined semiflexible polymers. Indeed, biopolymers' functionality is ruled by their conformation, which in turn is considerably modified in the geometrically confined or crowded environment inside the cell \cite{Fal-Koster2008, Fal-Reisner2005, Fal-Benkova2017}.

A common well-known theoretical framework used to describe the fundamental properties of a semiflexible polymer is the well-known worm-like chain model (WLC), which pictures a polymer as a thin wire with a flexibility given by its bending rigidity constant $\alpha$ \cite{PSaito-Saito1967}. The central quantity in this model is the persistence length defined by $2\alpha/(k_{B}T(d-1))$ \cite{Kleinert_2006, benetatos}, being $d$ the space dimension, however, here we simply use $\ell_{p}:=\alpha/(k_{B}T)$\footnote{For the sake of notation, it is hidden the dimension of the space in the persistence length definition. In those cases where an explicit dependence on the dimension is needed, it should be adequately scaled  by the factor $2/(d-1)$. }, which  is the characteristic length along the chain over which the directional correlation between segments disappears.  $k_{B}T$ is the thermal energy, with $k_{B}$ and $T$ the Boltzmann constant and $T$ the bath temperature, respectively. 
 \cite{19955-1}.
%\textcolor{red}{For the sake of notation, we hide the dimension of the space in the persistence length definition. In those cases where an explicit dependence on the dimensionality has needed, it should be adequately scaled by the factor...}

In the absence of thermal fluctuations,  when $\alpha\gg k_{B}T$, the conformations of the polymer are well understood through different curve configurations determined by variational principles \cite{Psim-Guven2012, Polcla-Guven2014}.
For the WLC model, the bending energy functional is given by
\begin{equation}
H[\mathbf{R}]=\frac{\alpha}{2}\int ds\kappa^2(s),
\end{equation}
where ${\bf R}(s)$ is a polymer configuration and  $\kappa(s)$ is the curvature of the chain, with $s$ the arc-length parameter.
Additional terms can be added to the Hamiltonian to account for other effects including multibody interactions, external fields, and constraints on the chain dimensions \cite{SW2005, Polcon-Chen2016}. 
When the thermal fluctuations are relevant $\alpha\simeq k_{B}T$, it is usual to introduce a statistical mechanics description. 
Since  $H[\mathbf{R}]$ represents the bending energy for a curve configuration ${\bf R}$, the most natural approach is to defined the canonical probability density
\begin{eqnarray}
\mathcal{P}\left(\boldsymbol{R}\right)\mathcal{D}\boldsymbol{R}:=\frac{1}{\mathcal{Z}_{\rm c}}\exp\left(-\frac{\ell_{p}}{2}\int ds \kappa^2\left(s\right)\right)\mathcal{D}\boldsymbol{R},
\label{ProbDensityCanonical}
\end{eqnarray}
where $\mathcal{Z}_{\rm c}$ is  the canonical partition function, and  $\mathcal{D}{\bf R}$ is an appropriate functional measure. In this description, the theory turns out to be a one-dimensional statistical field theory. 
Nonetheless, the theory is not easy  to tackle since $\kappa\left(s\right)$ acquires non-linear terms in ${\bf R}$. 
To avoid this difficulty, a different perspective was introduced by Saito's et al. \cite{PSaito-Saito1967}, where it was studied the following probability density function
\begin{eqnarray}
\mathcal{P}\left(\boldsymbol{T}\right)\mathcal{D}\boldsymbol{T}:=\frac{1}{\mathcal{Z}_{\rm s}}\exp\left(-\frac{\ell_{p}}{2}\int ds \kappa^2\left(s\right)\right)\mathcal{D}\boldsymbol{T},
\label{ProbDensityCanonicalSaito}
\end{eqnarray}
instead of Eq.~\eqref{ProbDensityCanonical}. Here $\mathcal{Z}_{\rm s}$ is the Saito's partition function and  $\mathcal{D}{\bf T}$ is an appropriate functional measure for the tangent direction of a given polymer configuration ${\bf R}$. The Saito's partition function can be  solved since one has  $\kappa^{2}(s)=\left({d{\bf T}(s)/ds}\right)^{2}$, thus one can relate $\mathcal{Z}_{\rm s}$ with the Feynman's partition function for a quantum particle in the spherical surface described by ${\bf T}^{2}=1$.  For the cases when the semiflexible polymer is in an open Euclidean space, the Saito's approach works very well. For instance, it reproduces the standard results of Kratky-Porod \cite{PSaito-Kratky1949}, among other results \cite{PSaito-Saito1967}. However, for the cases when the semiflexible polymer is confined to a bounded region of the space the Saito's approach is difficult to use, with some exceptional cases like the situation for semiflexible polymers confined to a spherical shell \cite{SW2005}. 

For semiflexible polymers in plane space, an alternative theoretical approach to the above formalisms  was introduced in \cite{Castro-JE}. This consists of postulating that each conformational realization of any polymer in the plane is described by a stochastic path satisfying the stochastic Frenet equations, defined by
$\frac{d}{ds}\mathbf{R}(s)=\mathbf{T}(s)$,  and
$\frac{d}{ds} \mathbf{T}(s)=  \kappa(s) {\bf N}\left(s\right)$,
where $\mathbf{R}(s)$ is the configuration  of the polymer, $\mathbf{T}(s)$ is the tangent vector to the curve describing the chain at $s$, $ {\bf N}(s):=\mathbf{\epsilon} \mathbf{T}(s)$ is the normal stochastic unit vector, with  $\mathbf{\epsilon}$ a rotation by an angle of $\pi/2$, and $\kappa(s)$ is the stochastic curvature that satisfies the following probability density function
\begin{eqnarray}
\mathcal{P}\left(\kappa\right)\mathcal{D}\kappa:=\frac{1}{\mathcal{Z}_{\rm s-c}}\exp\left(-\frac{\ell_{p}}{2}\int ds \kappa^2\left(s\right)\right)\mathcal{D}\kappa,
\label{ProbDensityCanonicalSTochastic}
\end{eqnarray}
where $\mathcal{Z}_{\rm s-c}$ is the partition function in the stochastic curvature formalism, and $\mathcal{D}\kappa$ is an appropriate measure for the curvature. This in particular, implies a white noise-like structure, i.e., $\langle \kappa(s) \rangle=0$ and $\langle \kappa(s) \kappa(s')\rangle=\delta(s-s')/\ell_p $ \cite{Castro-JE}.
This theoretical framework successfully explains, by first principles, the Kratky-Porod results for free chains confined into an open 2D-plane. Moreover, it correctly describes the mean square end-to-end distance for semiflexible polymers confined to a square box, a key descriptor of the statistical behavior of a polymer chain.

In the present work we carry out an extension of the stochastic curvature approach for semiflexible polymers in the three-dimensional space $\mathbb{R}^{3}$.  In particular, we analyze  the conformational states of a semiflexible polymer enclosed in a bounded region in three-dimensional space. This polymer is in a thermal bath with a uniform temperature. The shapes adopted by the polymer are studied through the mean-square end-to-end distance  as a function of the polymer total length  as well as its persistence length. In particular, we analyze the cases of a polymer confined  to a cube of side $a$ and a sphere of radius $R$. 

The plan of this paper is as follows. In Sect.~\ref{sec:preliminary}, we introduce the stochastic Frenet equations for the semiflexible polymers in three-dimensional spaces, and by using standard procedure we derive a corresponding Fokker-Planck equation. In particular, the Kratky-Porod result for polymers in a 3d open space is obtained. Sect. \ref{sec:compact} contains the derivation of the mean square end-to-end distance for semiflexible polymers confined to a compact domain.
In Sect. \ref{sec:results}, we present the analysis of the mean square end-to-end distance for the cases when the compact domain correspond with a cube of side $a$, and a sphere of radius $R$.
Finally, Sect. \ref{sec:conclusions} contains our concluding remarks.

%\newpage

\section{Preliminary notation and semiflexible polymers in 3D}\label{sec:preliminary}

Let us consider a polymer in a three-dimensional Euclidean space $\mathbb{R}^{3}$ as a space curve $\gamma$, ${\bf R}:I\subset \mathbb{R}\to\mathbb{R}^{3}$, parametrized by an arc-length, $s$. For each point $s\in I$, a Frenet-Serret trihedron can be defined in terms of the vector basis $\{{\bf T}(s), {\bf N}(s), {\bf B}(s)\}$, where ${\bf T}(s)=d{\bf R}/ds$ is the tangent vector, whereas ${\bf N}(s)$ and ${\bf B}(s)$ are the normal and bi-normal vector, respectively.  It is well known that each regular curve $\gamma$ satisfies the Frenet-Serret structure equations, namely,  $d{\bf T}/ds=\kappa(s){\bf N}$, $d{\bf N}/ds=-\kappa(s) {\bf T}-\tau(s) {\bf B}$ and $d{\bf B}/ds=\tau(s){\bf N}$, where $\kappa(s)$ and $\tau(s)$
 are the curvature and the torsion of the space curve. In addition, the fundamental theorem of space curves estates that given continuous functions $\kappa(s)$ and $\tau(s)$ one can determine the shape curve uniquely, up to a Euclidean rigid motion \cite{Fal-Montiel2009}.

\subsection{Stochastic curvature approach in 3D}
In order to study the conformational states of a semiflexible polymer,  we adapt the stochastic curvature approach introduced in \cite{Castro-JE} to the case of semiflexible polymers in 3D Euclidean space. For the 2D Euclidean space, the formalism starts by postulating that each conformational realization of any polymer is described by a stochastic path satisfying the stochastic Frenet equations. In the 3D case, it is enough to consider the following stochastic equations
\begin{subequations}\label{stoch-eq}
\begin{eqnarray}
\frac{d}{ds}{\bf R}(s)&=&{\bf T}(s),\label{stoch-eq0}\\
\frac{d}{ds}{\bf T}(s)&=&\mathbb{P}_{\bf T}\boldsymbol{\kappa}(s),
\end{eqnarray}
\end{subequations}
where ${\bf R}(s)$, ${\bf T}(s)$ and $\boldsymbol{\kappa}(s)$ are now random variables. $\boldsymbol{\kappa}(s)$ is named here stochastic vectorial curvature. Also, it has been introduced a normal projection operator $\mathbb{P}_{\bf T}=\boldsymbol{1}-{\bf T}\otimes {\bf T}$ such that ${\bf T}(s)\cdot \frac{d}{ds}{\bf T}(s) =0$. According to these equations, it can be shown that $\left|{\bf T}(s)\right|$ is a constant that can be fixed to unit, where $\left|~~\right|$ is the standard 3D Euclidean norm. The remaining geometrical notions also turn into  random variables as follows. The stochastic curvature is defined by $\kappa(s):=\left|\boldsymbol{\kappa}(s)\right|$. The stochastic normal and bi-normal vectors are defined by  ${\bf N}(s):=\boldsymbol{\kappa}(s)/\kappa(s)$ and  ${\bf B}(s):={\bf T}(s)\times \boldsymbol{\kappa}(s)/\kappa(s)$, respectively, where $\kappa(s)$ is the stochastic curvature. In addition,  the stochastic torsion is defined with the equation $\tau(s):={\bf N}(s)\cdot \frac{d}{ds}{\bf B}(s)$. 

In addition to the stochastic equations (\ref{stoch-eq}), the random variable $\boldsymbol{\kappa}(s)$ is distributed according to the probability density function 
\begin{eqnarray}
\mathcal{P}\left(\boldsymbol{\kappa}\right)\mathcal{D}\boldsymbol{\kappa}:=\frac{1}{\mathcal{Z}_{\rm s-c}}\exp\left(-\beta H\left[{\boldsymbol{\kappa}}\right]\right)\mathcal{D}\boldsymbol{\kappa},
\label{ProbDensity}
\end{eqnarray}
where  $H\left[\boldsymbol{\kappa}\right]=\frac{\alpha}{2}\int \boldsymbol{\kappa}^2 ds$ is the bending energy, and $\alpha$ is the bending rigidity modulus. This energy functional corresponds to the continuous form of the WLC model \cite{PSaito-Saito1967}. 
Also, in Eq. (\ref{ProbDensity}) $\mathcal{Z}_{\rm s-c}$ is an appropriate normalization constant, $\mathcal{D}{\boldsymbol{\kappa}}:=\prod_{i=1}^{3}\mathcal{D}\kappa_{i}$ is a functional measure, and $\beta=1/k_{B}T$ is the inverse of the thermal energy.  The Gaussian structure of the probability density  implies the zero mean $\left<\kappa_{i}(s)\right>=0$, and following $3D$ fluctuation theorem 
\begin{eqnarray}
\left<\kappa_{i}(s)\kappa_{j}(s^{\prime})\right>=\frac{1}{\ell_{p}}\delta_{ij}\delta(s-s^{\prime}),
\end{eqnarray}
where $\kappa_{i}(s)$ is the $i-$th component of the stochastic vectorial curvature $\boldsymbol{\kappa}(s)$.

\subsection{From Frenet-Serret stochastic equations to Hermans-Ullman equation in 3D}
In this section, we present the Fokker-Planck formalism corresponding to the stochastic equations (\ref{stoch-eq}). This description allows us to determine an equation for the probability density function associated to the position and direction of the endings of the polymer
$P\left({\bf R}, {\bf T}\left.\right|{\bf R}^{\prime}, {\bf T}^{\prime}; s\right)=\left<\delta\left({\bf R}-{\bf R}(s)\right)\delta\left({\bf T}-{\bf T}(s)\right)\right>$, where ${\bf R}$ and ${\bf R}^{\prime}$ are the ending positions of the polymer, and ${\bf T}$ and ${\bf T}^{\prime}$ are the corresponding directions. The parameter $s$ is the polymer length.

Now, the stochastic Frenet-Serret equations (\ref{stoch-eq}) can be identified with a multi-dimensional stochastic differential equation in the Stratonovich perspective, thus applying the standard procedure \cite{Fal-Gardiner1986},  we find the following Fokker-Planck type equation
\begin{eqnarray}
\frac{\partial P}{\partial s}+\nabla\cdot\left({\bf T}~P\right)=\frac{1}{2\ell_{p}}\Delta_{g}P\label{FPeq2},
\end{eqnarray}
where ${\bf T}$ is identified with the unit normal vector on  $S^{2}$, thus satisfies the  condition  ${\bf T}^{2}=1$. The operator $\Delta_{g}$ is the Laplace-Beltrami of the sphere $S^{2}$. Similarly, as the situation for semiflexible polymers confined to a plane space \cite{Castro-JE}, this equation is exactly the same as the one obtained by Hermans and Ullman in 1952 \cite{PSaito-Hermans1952}, where the heuristic parameter they included can now be identify exactly with $1/(2\ell_{p})$. In addition, we can make a contact with the Saito's approach \cite{PSaito-Saito1967} by considering the marginal probability density function
\begin{eqnarray}
\mathcal{Z}_{\rm s}\left({\bf T}, {\bf T}^{\prime}, s\right)\propto\int d^{3}{\bf R}d^{3}{\bf R}^{\prime} P\left({\bf R}, {\bf T}\left.\right|{\bf R}^{\prime}, {\bf T}^{\prime}, s\right).
\end{eqnarray}
Using the Hermans-Ullman equation, we can show that $\mathcal{Z}_{\rm c}$ satisfies a diffusion equation on a spherical surface with diffusion coefficient equal to $1/(2\ell_{p})$ \cite{PSaito-Saito1967}, that is, 
\begin{eqnarray}
\frac{\partial \mathcal{Z}_{\rm c}}{\partial s}=\frac{1}{2\ell_{p}}\Delta_{S^{2}}\mathcal{Z}_{\rm s}.
\end{eqnarray}
An immediate consequence of the above equation is the exponential decay of the 
correlation function between the two ending directions $C(L):=\left<{\bf T}(L)\cdot{\bf T}(0)\right>=\exp\left(-L/\ell_{p}\right)$, where $L$ is the polymer length. Indeed, this expectation value satisfies the following equation
$\frac{d}{ds}C(s)=\frac{1}{2\ell_{p}}\frac{1}{4\pi}\int_{S^{2}}d\Omega \left({\bf T}(s)\cdot{\bf T}(s^{\prime})\right)\Delta_{S^{2}}\mathbb{Z}$,
where $d\Omega$ is the solid angle and $4\pi$ is a normalization constant. Now,  we can integrate twice by parts the r.h.s. of last equation  and since $S^{2}$ is a compact manifold the boundary terms vanish. Also, using $\Delta_{S^{2}}{\bf T}=-\frac{2}{R^{2}}{\bf T}$, it is found that the correlation function satisfies the ordinary differential equation $\frac{d}{ds}C(s)=-\frac{2}{R^2}C(s)$. Now, we solve this equation using the initial condition $C(s^{\prime}=0)=1$, and the length of the polymer set up by $s=L$.

\subsection{Modified telegrapher equation }

As in the situation of the two-dimensional case \cite{Castro-JE},  we carry out a multipolar decomposition for HU equation in 3D. This consists of expanding  the probability density function $P\left({\bf R}, {\bf T}\left.\right|{\bf R}^{\prime}, {\bf T}^{\prime}; s\right)$ in a linear combination of the cartesian tensor basis elements $1$, $T_{i}$, $T_{i}T_{j}-\frac{1}{3}\delta_{ij}$, $T_{i}T_{j}T_{k}-\frac{1}{5}\delta_{\left(ij\right.}T_{\left.k\right)}$, $\cdots$,  where the symbols $(ijk)$ means symetrization of the indices $i, j, k$, that is, $\delta_{\left(ij\right.}T_{\left.k\right)}=\delta_{ij}T_{k}+\delta_{jk}T_{i}+\delta_{ki}T_{j}$ whose expansion coefficients  are hydrodynamic-like tensor fields. These tensors are $\rho({\bf R}, s)$, meaning by the manner how the ending positions are distributed in the space; $
\mathbb{P}\left({\bf R}, s\right)$, meaning as the local average of the polymer direction; $\mathbb{Q}_{ij}({\bf R}, s)$, pointing the way how the directions are correlated along the points of the space,  etc. These tensors are the moments associated to the cartesian tensor basis, {\it e.g.} $\mathbb{P}_{i}=\int \frac{d\Omega}{4\pi}T_{i}P\left({\bf R},{\bf T}, s\right)$. These fields  satisfy the following hierarchy equations 
\begin{eqnarray}
\frac{\partial \rho({\bf R}, s)}{\partial s}&=&-\partial_{i}\mathbb{P}^{i}\left({\bf R}, s\right),\label{1}\\
\frac{\partial \mathbb{P}_{i}({\bf R}, s)}{\partial s}&=&-\frac{1}{\ell_{p}}\, \mathbb{P}_{i}({\bf R}, s)-\frac{1}{3}\partial_{i}\rho({\bf R},s)-\partial^{j}\mathbb{Q}_{ij}\left({\bf R}, s\right),\label{2}\\
\frac{\partial \mathbb{Q}_{ij}\left({\bf R}, s\right)}{\partial s}&=&-\frac{3}{\ell_{p}} \mathbb{Q}_{ij}\left({\bf R}, s\right)-\frac{1}{5}\mathbb{T}_{ij}\left({\bf R}, s\right)-\partial^{k}\mathbb{R}_{ijk}\left({\bf R}, s\right),\label{3}\end{eqnarray}
where $\mathbb{T}^{ij}=\partial^{i}\mathbb{P}^{j}+\partial^{j}\mathbb{P}^{i}-\frac{2\delta^{ij}}{3}\partial_{k}\mathbb{P}^{k}$.  

Now, by combining Eqs. (\ref{1}) and (\ref{2})  we can obtain a modified telegrapher equation 
\begin{eqnarray}
\frac{\partial^2\rho\left({\bf R}, s\right)}{\partial s^2}+\frac{1}{\ell_{p}}\frac{\partial \rho\left({\bf R}, s\right)}{\partial s}=\frac{1}{3}\nabla^2\rho\left({\bf R}, s\right)+\partial_{i}\partial_{j}\mathbb{Q}^{ij}\left({\bf R}, s\right),\label{modTel}
\end{eqnarray}
where $\nabla^{2}$ is the 3D Laplacian.  In a mean-field point of view one can consider the preceding equation as an equation for the probability density function $\rho\left({\bf R}, s\right)$ under the presence of a mean-field $\mathbb{Q}_{ij}\left({\bf R}, s\right)$. In particular, $\mathbb{Q}_{ij}\left({\bf R}, s\right)$ does not play any role for the mean-square end-to-end distance for a semiflexible polymer in the open Euclidean 3D space. Indeed, let us defined the end-to-end distance as $\delta{\bf R}:={\bf R}-{\bf R}^{\prime}$, thus the mean-square end-to-end distance is given by 
\begin{eqnarray}
\left<\delta{\bf R}^{2}\right>_{\mathcal{D}}\equiv \int_{\mathcal{D}\times \mathcal{D}}\rho\left(\left.{\bf R}\right|{\bf R}^{\prime}, s\right)\delta{\bf R}^{2}d^{3}{\bf R}d^{3}{\bf R}^{\prime}.
\end{eqnarray}
Now, we implement the same procedure used in \cite{Castro-JE} to calculate the mean-square end-to-end distance in the open three-dimensional space $\mathcal{D}= \mathbb{R}^{3}$, where it is use the modify telegrapher equation (\ref{modTel}) and  the traceless property of $\mathbb{Q}_{ij}\left({\bf R},s\right)$.  We can reproduce the standard Kratky-Porod \cite{PSaito-Kratky1949} result for a semiflexible polymer in the three-dimensional space \cite{PSaito-Hermans1952, PSaito-Kratky1949}
\begin{eqnarray}
\left<\delta{\bf R}^{2}\right>_{\mathbb{R}^{3}}=2\ell_{p}L
-2\ell_{p}^{2}\left(1-\exp\left(-\frac{L}{\ell_{p}}\right)\right),\end{eqnarray}
with the typical well-known asymptotics limits: diffusive regime $\left<\delta{\bf R}^{2}\right>\simeq 2\ell_{p}L$ for $L\gg \ell_{p}$, and  ballistic regime $\left<\delta{\bf R}^{2}\right>\simeq L^2$ for $L\ll \ell_{p}$.

\section{Semiflexible polymer in a compact domain}\label{sec:compact}

In this section, we apply the hierarchy equations developed in the previous section in order to the determine the conformational states of a semiflexible polymer confined to a compact volume domain of size $V$.  From the hierarchy Eqs. (\ref{2}) and (\ref{3}), the tensors $\mathbb{P}_{i}({\bf R}, s)$ and $\mathbb{Q}_{ij}({\bf R}, s)$ damp out as $e^{-L/\ell_{p}}$ and $e^{-3L/\ell_{p}}$, respectively. Furthermore,  if we consider that the semiflexible polymer is enclosed in a compact volume $V:=\ell^3$, with a typical length $\ell$, thus as long as we consider cases when $3\ell/\ell_{p}$ is far from 1,  we may assume that $\mathbb{Q}_{ij}({\bf R}, s)$ is uniformly distributed. This condition corresponds to truncate the hierarchy equations at the second level, that is,  the only equations that survive in this approximation are  Eqs. (\ref{1}) and (\ref{2}).  

In the latter situation, the distribution, $\rho({\bf R}, s)$ of the endings of the semiflexible polymer is describes through the following telegrapher's equation 
 \begin{eqnarray}
\frac{\partial^2\rho\left({\bf R}, s\right)}{\partial s^2}+\frac{1}{\ell_{p}}\frac{\partial \rho\left({\bf R}, s\right)}{\partial s}=\frac{1}{3}\nabla^2\rho\left({\bf R}, s\right), 
\label{tel3d}\end{eqnarray}
that satisfies the initial conditions 
\begin{eqnarray}
\lim_{s\to 0}\rho\left({\bf R}\right.\left|{\bf R}^{\prime}, s\right)&=&\delta^{(3)}\left({\bf R}-{\bf R}^{\prime}\right),\label{cond1}\\
\lim_{s\to 0}\frac{\partial\rho\left({\bf R}\right.\left|{\bf R}^{\prime}, s\right)}{\partial s}&=&0.\label{cond2}
\end{eqnarray}
The condition (\ref{cond1})  means that the polymers' ends coincide when the polymer length is zero, whereas (\ref{cond2}) means that the polymer length does not change spontaneously. In addition, since the polymer is enclosed in the compact domain $\mathcal{D}$ of volume $V(\mathcal{D})$, we also impose a Neumann boundary condition 
\begin{eqnarray}
\left.\nabla\rho\left({\bf R}\right.\left|{\bf R}^{\prime}, s\right)\right|_{{\bf R}, {\bf R}^{\prime}\in \partial D}=0, ~~~\forall s,
\end{eqnarray}
where $\partial D$ is  a surface bounding the domain $\mathcal{D}$. This boundary condition means that the polymer does not cross the boundary neither wrap the domain.  The procedure to obtain a solution of the above telegrapher's equation (\ref{tel3d}) is identically to the one developed in \cite{Castro-JE}. We just have to take into account the right factors and the dimensionality considerations. In this sense, the probability density function is given by
\begin{eqnarray}
\rho\left(\left.{\bf R}\right|{\bf R}^{\prime};s\right)=\frac{1}{V({\mathcal{D}})}\sum_{{\bf k}\in I}G\left(\frac{s}{2\ell_{p}}, \frac{4\ell^{2}_{p}}{3}\lambda_{{\bf k}}\right)\psi^{\dagger}_{{\bf k}}\left({\bf R}\right)\psi_{{\bf k}}\left({\bf R}^{\prime}\right)\label{density}
\end{eqnarray}
where we recall from \cite{Castro-JE}
\begin{eqnarray}
G(v,w)=e^{-v}\left[\cosh\left(v\sqrt{1-w}\right)+\frac{\sinh\left(v\sqrt{1-w}\right)}{\sqrt{1-w}}\right],
\end{eqnarray}
and $\{\psi_{\bf k}\}$ and $\{\lambda_{{\bf k}}\}$ are a complete set of orthonormal eigenfunctions and a set of corresponding eigenvalues of the Laplace operator $-\nabla^{2}$ in $\mathbb{R}^{3}$. Notice that each $\psi_{\bf k}({\bf R})$ must satisfy the Neumann boundary equation $\left.\nabla \psi_{{\bf k}}\right|_{{\bf R}\in \partial \mathcal{D}}=0$. In addition, it is known \cite{Fal-Feshbach1953, Fal-Chavel1984} that for Neumann boundary Laplacian eigenvalue problem  there is a zero eigenvalue $\lambda_{0}=0$ corresponding to a positive eigenfunction given by $\psi_{0}={1/\sqrt{V}}$.

Now, using (\ref{density}), the mean-square end-to-end distance $\left<\delta{\bf R}^{2}\right>_{\mathcal{D}}$ can be computed in the standard fashion by\begin{eqnarray}
\left<\left(\delta {\bf R}\right)^{2}\right>_{\mathcal{D}}=\sum_{{\bf k}\in I}a_{k}G\left(\frac{s}{2\ell_{p}}, \frac{4\ell^{2}_{p}}{3}\lambda_{{\bf k}}\right),
\end{eqnarray}
where the coefficients $a_{k}$ are obtained from
\begin{eqnarray}
a_{k}=\frac{1}{V(\mathcal{D})}\int_{\mathcal{D}\times \mathcal{D}}\left({\bf R}-{\bf R}^{\prime}\right)^{2}\psi^{\dagger}_{{\bf k}}({\bf R})\psi_{{\bf k}}({\bf R}^{\prime})d^{3}{\bf R}d^{3}{\bf R}^{\prime}.
\end{eqnarray}
We can have a further simplification since after squaring the end-to-end distance inside the last integral. It is not difficult to see that the square terms ${\bf R}^{2}$ and ${\bf R}^{\prime 2}$ in $({\bf R}-{\bf R}^{\prime})^{2}$ only the zero mode contribute, thus we have
\begin{eqnarray}
\left<\left(\delta{\bf R}\right)^{2}\right>_{\mathcal{D}}=2\sigma^2\left({\bf R}\right)-\frac{2}{V\left(\mathcal{D}\right)}\sum_{{\bf k}\neq 0}{\bf r}^{*}_{{\bf k}}\cdot {\bf r}_{{\bf k}}~G\left(\frac{s}{2\ell_{p}}, \frac{4\ell^{2}_{p}}{3}\lambda_{{\bf k}}\right),\label{MSD}
\end{eqnarray}
where $\sigma^2\left({\bf R}\right):=\left<{\bf R}^{2}\right>_{g}-\left<{\bf R}\right>^{2}_{g}$ is called mean-square end position,  $\left<\cdots\right>_{g}:=\frac{1}{V\left(\mathcal{D}\right)}\int_{\mathcal{D}}d^{3}{\bf R}\cdots$ is termed geometric average, and the factor  ${\bf r}_{{\bf k}}:=\int_{\mathcal{D}}{\bf R}\psi_{{\bf k}}\left({\bf R}\right)d^{3}{\bf R}$ for ${\bf k}\neq 0$. The factor ${\bf r}_{{\bf k}}$ can be written in a simpler form for Neumann boundary conditions, since $\psi_{{\bf k}}=-\frac{1}{\lambda_{\bf k}}\nabla^{2}\psi_{{\bf k}}$ and by integrating out by parts, this factor is expressed in terms of a boundary integral 
\begin{eqnarray}
{\bf r}_{{\bf k}}=\frac{1}{\lambda_{\bf k}}\oint_{\partial \mathcal{D}}dS~{\bf n}\psi_{\bf k}\left({\bf R}_{S}\right), \label{expresion}
\end{eqnarray}
where ${\bf R}_{S}\in \partial\mathcal{D}$ and $dS$ is the area element of $\partial \mathcal{D}$. Since the function $G\left(v, w\right)$ decays exponentially as the polymer length gets larger values, we can convince ourselves that twice the mean-square end position corresponds to a saturation value for the mean-square end-to-end distance. An additional property of ${\bf r}_{\bf k}$ is the identity
\begin{eqnarray}
\frac{1}{V\left(\mathcal{D}\right)}\sum_{{\bf k}\neq 0}{\bf r}_{k}^{*}{\bf r}_{\bf k}=\sigma^{2}\left({\bf R}\right).\label{equidentity}
\end{eqnarray}
 This identity can be proved using the completeness relation of the eigenfunctions, that is,  $\sum_{\bf k}\psi^{*}_{\bf k}\left({\bf R}\right)\psi_{\bf k}\left({\bf R}^{\prime}\right)=\delta^{(3)}\left({\bf R}^{\prime}-{\bf R}\right)$. This identity allows us to proved that in general $\left<\left(\delta{\bf R}\right)^{2}\right>_{\mathcal{D}}$ starts at zero.

\section{Results}\label{sec:results}

\subsection{Semiflexible polymer enclosed by a cube surface}
In this section we provide results for the mean-square end-to-end distance for a semiflexible polymer enclosed inside of a cube domain. All the problem is reduced to solve the Neumann eigenvalue problem $-\nabla^2\psi=\lambda\psi$ with Neumann boundary condition when the compact domain is $\mathcal{C}:=\left\{(x,y,z)\in \mathbb{R}^{3}: 0\leq x\leq a, 0\leq y\leq a, 0\leq z\leq a \right\}$ is a cube  of side $a$ in the positive octant. This problem is widely studied in different mathematical physics problems \cite{Fal-Feshbach1953, Grebenkov}. The eigenfunctions in this case can be given by 
\begin{eqnarray}
\psi_{\bf k}\left({\bf R}\right)=\frac{N_{nmp}}{a^{3/2}}\cos\left(\frac{\pi n}{a}x\right)\cos\left(\frac{\pi m}{a}y\right)\cos\left(\frac{\pi p}{a}z\right),
\end{eqnarray}
where $x, y$ and $z$ are the standard  cartesian coordinates, and ${\bf R}=(x, y,z)$ is the usual vector position. 
The eigenfunctions are enumerated by the collective index $n m p$, with $n, m, p=0, 1,2, \cdots$.
$N_{nmp}$ is a normalization constant with respect to the volume of the cube $V(\mathcal{D})=a^{3}$, whose values are given by $N_{000}=1$; $N_{n00}=N_{0n0}=N_{00n}=\sqrt{2}$, for $n\neq 0$; $N_{np0}=N_{n0p}=N_{0np}=2$, for $n, p\neq 0$; and $
N_{npm}=2\sqrt{2}$, for $n,m,p\neq 0$. The eigenvalues of the Laplacian are given by $\lambda_{{\bf k}}={\bf k}^2$, where ${\bf k}=\left(\frac{\pi n}{a}, \frac{\pi m}{a}, \frac{\pi p}{a}\right)$. 
 Now, we proceed to calculate ${\bf r}_{{\bf k}}$ using its definition, that is, ${\bf r}_{{\bf k}}=\int_{\mathcal{C}}{\bf R}\psi_{\bf k}\left({\bf R}\right)d^{3}{\bf R}$. The three components are given by
 \begin{eqnarray}
\left( {\bf r}_{\bf k}\right)_{x}&=&-\sqrt{2}\frac{a^{5/2}}{n^2\pi^2}\left(1-(-1)^{n}\right)\delta_{m0}\delta_{p0}\nonumber\\
\left( {\bf r}_{\bf k}\right)_{y}&=&-\sqrt{2}\frac{a^{5/2}}{m^2\pi^2}\left(1-(-1)^{m}\right)\delta_{n0}\delta_{p0}\nonumber\\
\left( {\bf r}_{\bf k}\right)_{z}&=&-\sqrt{2}\frac{a^{5/2}}{p^2\pi^2}\left(1-(-1)^{p}\right)\delta_{n0}\delta_{m0}
 \end{eqnarray}
 
\begin{figure}[h!]
\begin{center}
\includegraphics[scale=1.3]{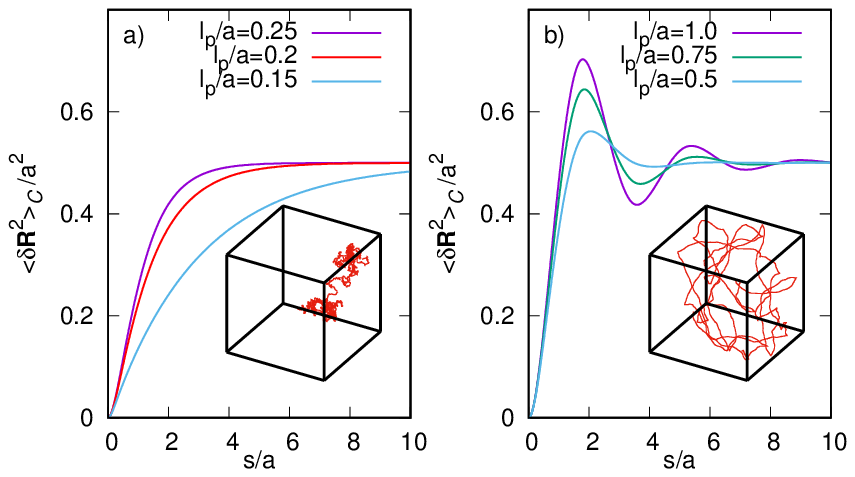}
\caption{\small %\textcolor{red}{Mean-square end-to-end distance for polymers confined in the volume enclosed by a cube as a function of the persistence length. Note that $\left<\delta{\bf R}^{2}\right>_{\mathcal{C}}$ shows an oscillating behavior for values of persistence length satisfying the relation $\ell_{p}/a > \sqrt{3}/(2\pi)$.}
Monotonous and oscillating behaviors of the mean square end-to-end distance (Eq. (\ref{MSDCube})) of polymers  with $\ell_{p}$ below [a)] and above [b)] the critical persistence length $\ell^{*}_{p}=\sqrt{3}/(2\pi)a$ in cubic confinement. Inside the plotting area we sketch the conformational states of each class of polymers.
}
\label{fig1}
\end{center}
\end{figure}

In the following, we use the general expression (\ref{MSD}) for the mean-square end-to-end distance. The mean-square end position can be easily calculated $\sigma^{2}\left({\bf R}\right)=\frac{a^{2}}{4}$. Since the Kronecker deltas in ${\bf r}_{\bf k}$ each contribution of $\left({\bf r}_{\bf k}\right)_{i}$ is the same, thus taking into account the correct counting factor the mean square end-to-end distance is
\begin{eqnarray}
\left<\delta{\bf R}^{2}\right>_{\mathcal{C}}=\frac{a^{2}}{2}-24a^2\sum_{k=1}^{\infty}\frac{\left(1-(-1)^{k}\right)}{k^{4}\pi^{4}}G\left(\frac{s}{2\ell_{p}}, \frac{4}{3}\left(\frac{\ell_{p}}{a}\right)^{2}\pi^{2}k^{2}\right). \label{MSDCube}
\end{eqnarray}
Following, the same line of argument performed in \cite{Castro-JE}, it is observed that $24\sum_{k=1}^{\infty}\frac{\left(1-(-1)^{k}\right)}{k^{4}\pi^{4}}=\frac{1}{2}$ consistently with (\ref{equidentity}), thus up to a  numerical error of $10^{-2}$, we claim that
{\begin{eqnarray}
\frac{\left<\delta{\bf R}^2\right>_{\mathcal{C}}}{a^2}&\simeq&\frac{1}{2}-\frac{1}{2}\exp\left(-\frac{L}{2\ell_{p}}\right)
\left\{ \cosh\left[\frac{L}{2\ell_{p}}\left(1-\frac{4\pi^2}{3}\frac{\ell^2_{p}}{a^2}\right)^{\frac{1}{2}}\right]\right.\nonumber\\&+&\left.\left(1-\frac{4\pi^2}{3}\frac{\ell^2_{p}}{a^2}\right)^{-\frac{1}{2}}\sinh\left[\frac{L}{2\ell_{p}}\left(1-\frac{4\pi^2}{3}\frac{\ell^2_{p}}{a^2}\right)^{\frac{1}{2}}\right]\right\}.
\label{approxx}
\end{eqnarray}}
Let us remark that for any fixed value of $a$, the r.h.s of (\ref{approxx}), as a function of $L$, shows the existence of a critical persistence length, $\ell_{p}^{*}=\sqrt{3}a/(2\pi)$ such that for all values of $\ell_{p}>\ell_{p}^{*}$ it exhibits an oscillating behavior, whereas for $\ell_{p}<\ell_{p}^{*}$ it is monotonically increasing.  In Fig. (\ref{fig1}), we show the behavior of the mean-square end-to-end distance versus the length of the polymer for several values of the persistence length below and above $\ell_p^*$. Moreover, we also show sketches of conformational states corresponding to the monotonous and oscillating behaviors of the mean square end-to-end distance. In addition, it is noticeably that the same mathematical structure as the mean-square end-to-end distance found by Spakowitz and Wang \cite{SW2005} for semiflexible polymers wrapping a spherical shell, and recently for semiflexible polymers confined to a square box \cite{Castro-JE}.  

\subsection{Semiflexible polymer enclosed by a spherical surface}
In this section we provide results for the mean-square end-to-end distance for a semiflexible polymer enclosed inside of a spherical domain. All the problem is reduced to solve the Neumann eigenvalue problem $-\nabla^2\psi=\lambda\psi$ with Neumann boundary condition when the compact domain is $\mathcal{B}:=\left\{{\bf r}\in \mathbb{R}^{3}: {\bf r}^{2}\leq R^{2}\right\}$ is a center ball of radius $R$. This problem is widely studied in different mathematical physics problems \cite{Fal-Feshbach1953, Grebenkov}. The eigenfunctions in this case can be given in terms of spherical Bessel functions $j_{\ell}\left(x\right)$ and spherical harmonic functions $Y_{\ell m}\left(\theta, \varphi\right)$, 
\begin{eqnarray}
\psi_{\ell m k}\left(r, \theta, \varphi\right)=N_{\ell k}~j_{\ell}\left(\alpha_{\ell k}\frac{r}{R}\right)Y_{\ell m}\left(\theta, \varphi\right)
\end{eqnarray}
where $r, \theta$ and $\varphi$ are the standard  spherical coordinates. The factor $N_{\ell k}$ is a normalization constant with respect to the volume of the ball $\mathcal{B}$, given by 
\begin{eqnarray}
N_{\ell k}=\frac{\sqrt{2}}{R^{3/2}}\frac{\alpha_{\ell k}}{j_{\ell}\left(\alpha_{\ell k}\right)\left(\alpha_{\ell k}^{2}-\ell\left(\ell+1\right)\right)^{1/2}}.\label{norma}
\end{eqnarray}
The coefficients $\alpha_{\ell k}$ are the roots of $\partial j_{\ell}\left(x\right)/\partial x$, which by using the identity $\ell j_{\ell-1}\left(x\right)-\left(\ell+1\right)j_{\ell+1}=\left(2\ell+1\right)\partial j_{\ell}\left(x\right)/\partial x$, that satisfies the equation $\ell j_{\ell-1}\left(\alpha_{\ell k}\right)=\left(\ell+1\right)j_{\ell+1}\left(\alpha_{\ell k}\right)$. The eigenfunctions are enumerated by the collective index $\ell m k$, with $\ell=0, 1,2, \cdots$ countings the order of spherical Bessel functions, $m=-\ell, -\ell+1, \cdots, \ell$, and $k=1,2,3, \cdots$ counting zeros. The eigenvalues of the Laplacian are given by $\lambda_{\ell m k}={\alpha^{2}_{\ell k}/R^{2}}$, which are independent of the numbers $m$.  Now, we proceed to calculate ${\bf r}_{\ell m k}$ using (\ref{expresion}). It is enough to calculate $\oint_{S^{2}} dS~{\bf n}~Y_{\ell m}\left(\theta, \varphi\right)$, since ${\bf n}\propto Y_{1m}$, thus $\oint_{S^{2}} dS~{\bf n}~Y_{1, \pm1}\left(\theta, \varphi\right)=-\sqrt{\frac{2\pi}{3}}R^{2}\left(\pm 1, i, 0\right)$ and $\oint_{S^{2}} dS~{\bf n}~Y_{1, 0}\left(\theta, \varphi\right)=2\sqrt{\frac{\pi}{3}}R^2\left(0,0,1\right)$. Now, we call $\alpha_{1k}:=\alpha_{k}$, then using (\ref{norma}) one has
\begin{eqnarray}
{\bf r}_{1, \pm 1, k}&=&-2\sqrt{\frac{\pi}{3}}\frac{R^{5/2}}{\alpha_{k}\left(\alpha^{2}_{k}-2\right)^{1/2}}\left(\pm 1, i, 0\right),\\
{\bf r}_{1, 0, k}&=&2\sqrt{\frac{2\pi}{3}}\frac{R^{5/2}}{\alpha_{k}\left(\alpha^{2}_{k}-2\right)^{1/2}}\left(0, 0, 1\right),
\end{eqnarray}
where roots $\{\alpha_{k}\}$ satisfy the equation $j_{0}\left(\alpha_{k}\right)=2j_{2}\left(\alpha_{k}\right)$. Using explicit functions of the spherical Bessel functions the root condition is $F(\alpha_{k})=0$, where
\begin{eqnarray}
F\left(x\right)=\left(\frac{x^2}{2}-1\right)\sin x+x\cos x.\label{Raices}
\end{eqnarray}

In the following, we use the general expression (\ref{MSD}) for the mean-square end-to-end distance. We calculate the mean-square end position, $\sigma^{2}\left({\bf R}\right)=\frac{3}{5}R^2$, and use the factors ${\bf r}_{\ell m k}$, thus the mean square end-to-end distance is
\begin{eqnarray}
\left<\delta{\bf R}^{2}\right>_{\mathcal{B}}=\frac{6}{5}R^{2}-12R^2\sum_{k=1}^{\infty}\frac{1}{\alpha^2_{k}(\alpha_{k}^{2}-2)}G\left(\frac{s}{2\ell_{p}}, \frac{4}{3}\left(\frac{\ell_{p}}{R}\right)^{2}\alpha^{2}_{k}\right). \label{MSDSBall}
\end{eqnarray}
Following, the same line of argument performed in \cite{Castro-JE}, we observe numerically that
$12\sum_{k=1}^{N}\frac{1}{\alpha^{2}_{k}\left(\alpha_{k}^2-2\right)}\to6/5$ as $N$ increases, this is consistent with Eq. (\ref{equidentity}). Thus up to a numerical error $10^{-2}$, we claim that
{\begin{eqnarray}
\frac{\left<\delta{\bf R}^2\right>_{\mathcal{B}}}{R^2}&\simeq&\frac{6}{5}-\frac{6}{5}\exp\left(-\frac{L}{2\ell_{p}}\right)
\left\{ \cosh\left[\frac{L}{2\ell_{p}}\left(1-\frac{4\alpha_{1}^2}{3}\frac{\ell^2_{p}}{R^2}\right)^{\frac{1}{2}}\right]\right.\nonumber\\&+&\left.\left(1-\frac{4\alpha_{1}^2}{3}\frac{\ell^2_{p}}{R^2}\right)^{-\frac{1}{2}}\sinh\left[\frac{L}{2\ell_{p}}\left(1-\frac{4\alpha_{1}^2}{3}\frac{\ell^2_{p}}{R^2}\right)^{\frac{1}{2}}\right]\right\}.
\label{approx2}
\end{eqnarray}}

\begin{figure}[h!]
\begin{center}
\includegraphics[scale=1.3]{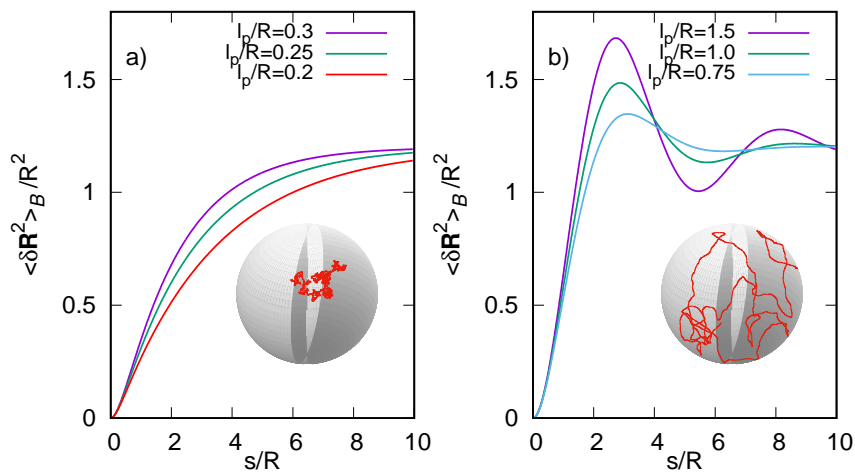}
\caption{\small %\textcolor{red}{Mean-square end-to-end distance for polymers confined in the volume enclosed by a sphere as a function of the persistence length. Note that $\left<\delta{\bf R}^{2}\right>_{\mathcal{D}}$ shows an oscillating behavior for values of persistence length satisfying the relation $\ell_{p}/R > 3/(4\alpha_{1})$.}
Monotonous and oscillating behaviors of the mean square end-to-end distance (Eq. (\ref{MSDSBall}))  of polymers  with $\ell_{p}$ below [a)] and above [b)] the critical persistence length $\ell^{*}_{p}=\sqrt{3}R/(2\alpha_1)$ in spherical confinement. Inside the plotting area we sketch the conformational states of each class of polymers.
}
\label{fig2}
\end{center}
\end{figure}

Let us remark that for any fixed value of $R$, the r.h.s of (\ref{approx2}), as a function of $L$, shows the existence of a critical persistence length, $\ell_{p}^{*}=\sqrt{3}R/(2\alpha_{1})$, with $\alpha_{1}\simeq 2.08158 $ according to (\ref{Raices}), such that for all values of $\ell_{p}>\ell_{p}^{*}$ it exhibits an oscillating behavior, whereas for $\ell_{p}<\ell_{p}^{*}$ it is monotonically increasing.  In Fig. (\ref{fig2}), we show the behavior of the mean-square end-to-end distance versus the length of the polymer for several values of the persistence length below and above $\ell_p^*$. Moreover, we also show sketches of conformational states corresponding to the monotonous and oscillating behaviors of the mean square end-to-end distance. In addition, it is noticeably that the same mathematical structure as the mean-square end-to-end distance found by Spakowitz and Wang \cite{SW2005} for semiflexible polymers wrapping a spherical shell, and recently for semiflexible polymers confined to a square box \cite{Castro-JE}.  

\section{Concluding remarks}\label{sec:conclusions}

In this work we carry out an extension of the stochastic curvature formalism introduced in \cite{Castro-JE} to analyze the conformational states of a semiflexible polymer in a thermal bath for the cases when the polymer is in the open space $ \mathbb{R}^ {3}$, and when is  in a bounded domain $ \mathcal{D}\subset \mathbb{R}^{3}$. The basic idea of formalism in the 3D case is followed by two postulates, that is, that each conformational state corresponds to the realization of a path described by the stochastic Frenet-Serret equations (\ref{stoch-eq}),  to introduce an stochastic curvature vector $\boldsymbol{k}\left(s\right)$, and a second postulate that gives the manner how $\boldsymbol{\kappa}(s)$ is distributed according to the thermal fluctuations.

In the case of a polymer in an open space $\mathbb{R}^3$, the standard Kratky-Porod formula for polymers is reproduced in three dimensions \cite{PSaito-Kratky1949}, while when the polymer is confined to a space bounded region $\mathcal{D}\subset\mathbb{R}^{3}$  the conformational states shows the existence of a critical persistence length $\ell_{p}^{*}$ such that for all values of $\ell_{p}>\ell_{p}^{*}$ the mean square distance from end-to-end exhibits an oscillating behavior, while for $\ell_{p} <\ell_{p}^{*}$ it exhibits a monotonic behavior in both cases of a cubic region  and a spherical region . Furthermore, for each value of $\ell_{p}$, the function converges to the twice of the mean-square end position $\sigma^{2}\left({\bf R}\right)$, that is, twice the variance of ${\bf R}^2$ with respect to the volume of the domain. 
The critical persistence length, therefore, distinguishes two conformational behaviors of the semiflexible polymer in the bound domain.  On one hand, polymers with persistence length below the critical value have a conformation similar to a Brownian random path. On the other hand, polymers with persistence length above the critical value adopt smooth conformations. 
In addition, it is highlighted  that the mean square end-to-end distance exhibits the same  mathematical form for the discussed cases along with the manuscript (see Eq.(31) and Eq. (38)), and with the results reported for a polymer enclosed to a square box and rolling up a spherical surface \cite{Castro-JE, SWPRL}. 
Nevertheless, the values-difference of saturation and the critical persistence length reflects the particular geometric nature of the compact domain, including the dimensionality of the space. 
Note the particular mathematical expression in our work is due to the probability density function of the polymer's ends which is governed by a modified telegrapher equation. 
%These results are of the same type as that found by Wang and Spakowitz \cite{SW2005} for a semiflexible polymer that envelops a spherical surface and recently for a square box in \cite{Castro-JE}. 
As a consequence of this resemblance, it can be concluded that the shape transition from oscillating to monotonous conformational states provides furthermore evidence of a universal signature for a semiflexible polymer enclosed in compact space.

\section*{Conflict of Interest Statement}

The authors declare that the research was conducted in the absence of any commercial or financial relationships that could be construed as a potential conflict of interest.

\section*{Author Contributions}

Both authors contributed to the formulation of the method, and the writing of the manuscript. JER contributed in the numerical analysis that provide the figures, while PCV in the mathematical calculations.

\section*{Funding}

\section*{Acknowledgments}

P.C.V. acknowledges financial support CONACyT. .

\bibliographystyle{ieeetr}
\bibliography{ref}

\end{document}